# The political economy of big data leaks:
# Uncovering the skeleton of tax evasion

Pier Luigi Sacco*
*DiSFiPEQ, University of Chieti-Pescara, Viale Pindaro 42, 65127 Pescara, Italy*
*metaLAB (at) Harvard, 42 Kirkland St, 02138 Cambridge MA*
pierluigi.sacco@unich.it, pierluigi_sacco@fas.harvard.edu

Alex Arenas
*Departament d'Enginyeria Informática i Matemátiques, Universitat Rovira I Virgili, Av.da Paisos Catalans, 26, 43007 Tarragona, Spain*
alexandre.arenas@urv.cat

Manlio De Domenico*
*Department of Physics, University of Padova, Via Francesco Marzolo 8, 35121 Padova, Italy*
manlio.dedomenico@unipd.it

* Corresponding authors

A.A. and M.D.D. acknowledge financial support by the Spanish government through grant FIS2015-38266. M.D.D. also acknowledges financial support from the Spanish program Juan de la Cierva (IJCI-2014-20225). A.A. also acknowledges support by Ministerio de Economía y Competitividad (Grants No. PGC2018-094754-B-C21 and No. FIS2015-71582-C2-1), Generalitat de Catalunya (Grant No. 2017SGR-896), Universitat Rovira i Virgili (Grant No. 2019PFR-URV-B2-41), ICREA Academia, and the James S. McDonnell Foundation (Grant No. 220020325).

**Competing interests**

The authors declare that they have no competing interests.

**Author's contributions**

The authors contributed equally to this work.

**Abstract**

After the leak of 11.5 million documents from the Panamanian corporation Mossack Fonseca, an intricate network of offshore business entities has been revealed. The emerging picture is that of legal entities, either individuals or companies, involved in offshore activities and transactions with several tax havens simultaneously which establish, indirectly, an effective network of countries acting on tax evasion. The analysis of this network quantitatively uncovers a strongly connected core (a rich-club) of countries whose indirect interactions, mediated by legal entities, form the skeleton for tax evasion worldwide. Intriguingly, the rich-club mainly consists of well-known tax havens such as British Virgin Islands and Hong Kong, and major global powers such as China, Russia, United Kingdom and United States of America. The analysis provides a new way to rank tax havens because of the role they play in this network, and the results call for an international coordination on taxation policies that take into account the complex interconnected structure of tax evaders in a globalized economy.

*1. Introduction*

Tax evasion is a major public policy issue (Zucman and Piletty, 2015), both in terms of fairness (it is concentrated among the rich; Alstadsæter et al., 2019), and of sustainability (it may undermine social compliance toward the provision of public goods; Kamm et al., 2021). Major global institutions have therefore committed themselves to eradicate it through coordinated action (Crasnic and Hakelberg, 2021). However, results have been so far not matching expectations and goals due to a number of countervailing forces (Cockfield, 2018), including more or less overt resistance by a number of countries which benefit from the current status quo (Zhou and Bagheri, 2021), including the USA (Brinson, 2019).

A major role in tax evasion and avoidance is played by tax havens, which offer an enabling institutional environment guaranteeing very low or null taxation of income and wealth, little concern for lawfulness of their origin, and plenty of legal tools to shield or conceal ownership (Gravelle, 2009). Despite that each tax haven has a particular history and political-economic rationale of its own (Ogle, 2017; Raposo and Mourão, 2013), and that a country's capacity to function as a tax haven may rise but also decline (Robertson, 2021), it would be a mistake to think of them as isolated entities, or that there is a binary separation between countries that should be considered tax havens and countries that should not (Cobham et al, 2015), so that any such classification has to be constantly checked and updated (Dharmapala and Hines, 2009). Tax havens only make sense as nodes of a global network, which serves a plurality of categories of economic agents: large multinational enterprises (Jones et al, 2018), criminal cartels (Cobham, 2012), ultrarich private individuals (Saez and Zucman, 2019), and so on. But for such network to be functional, it is mandatory that its structure and flows remain opaque and impossible to track (Janský et al, 2021). This secrecy reduces public awareness of the scale and seriousness of the problem and of the damage to the public interest, and even partial elimination of large tax havens would be welfare improving for both non-haven countries and for the remaining havens (Slemrod and Wilson, 2009).

For this reason, recent research on tax evasion makes an increasingly systematic use of big data (Cockfield et al, 2019) and social network analysis (Crofts and Sigler, 2018). More generally, computational social science may become an extremely useful interdisciplinary arena to improve the understanding and monitoring of global tax avoidance and tax evasion flows (Ferwerda et al, 2020). Of special interest in this regard are the breakdowns of secrecy that occur when data from the archives of legal firms operating in tax havens are leaked, providing a wealth of information on tax evasion networks. However, despite its objective value, such information is not easily exploitable due to the extreme complexity of the flows of activities and to the very high number of actors and entities involved. These difficulties provide an additional reason to appreciate the potential contribution of computational social science to the parsing of such a large, diverse and elusive amount of data.

Among such data leaks, one of the biggest and most important so far has been that of the so called "Panama papers" in 2016, deriving from a breach of secrecy of the Panama-based law firm Mossack Fonseca – a telling example of the role and importance of advanced business services firms in enabling tax evasion, but also as a potential informational gateway to unveil its structure and functioning (Poon et al, 2019).

Compared to previous leaks, the Panama Papers one is unprecedented in size and richness of information: it contained 2.6 TB of data, compared to the 3.3GB of the HSBC Files leak of 2015, 4.4 GB of the Luxembourg Tax Files of 2014, and to the 1.7 GB of the Wikileaks of 2010 (source: The Economist, April 9, 2016). It is therefore not surprising that the Panama Papers Data (PPD) have been thoroughly analyzed with computational techniques. Some contributions have analyzed how PPD could help to shed light upon the tax avoidance dynamics and flows of particular geopolitical areas, such as the Middle East (Rabab'Ah et al, 2016), China and Hong Kong (Michael and Goo, 2019), or Portugal (Barbosa et al, 2017). Other studies (Peacock and Weissinger, 2018; Srivastava and Singh, 2018) offer an analysis of some of the main structural characteristics of the global tax evasion network as revealed by the PPD and by the subsequent Paradise Papers leak occurred in 2017. Kejriwal and Dang (2020) build several higher-order networks from PPD highlighting the existence of relevant structural differences between tax evasion networks and typical information and social networks, whereas Garcia Alvarado and Mandel (2019) characterize the high level of structural coherence of the global tax evasion network as emerging from PPD and provide criteria to identify the tax havens that should be main targets to dismantle the network. Dominguez et al (2020) provide a detailed reconstruction of the flows and connectivity across the main offshoring countries and regions as emerging from PPD. Zhuhadar and Ciampa (2019) parse PPD by means of cognitive computing to highlight various important patterns, including the top intermediary countries, the distribution of offshore companies by country and the relative sizes of country-to-country flows; further semantic analysis in Zhuhadar and Ciampa (2021) unveils specialized subnetworks according to the nature of the ownership of offshore entities (personal vs. corporate).

Such wealth of data and insight has allowed to appreciate how the tax evasion network is not separated from the financial flows that link major world cities but rather functions as a sort of undercover counterpart to the visible transactions (Cloke and Brown, 2019). Moreover, global financial elites are characterized by complex patterns of coownership of offshore assets which can constrain the discretionary power of local dictatorships in the enforcement of property rights (Derpanopoulos, 2018). Secret offshore vehicles are used by hundreds of listed companies to fund corruption, avoid taxation, and even expropriate shareholders (O'Donovan et al, 2019). And finally, the tax evasion network functions as an all-purpose financial highway where legal and suspicious entities are intertwined (Joaristi et al, 2019).

Our paper adds to this growing literature by providing a simple yet very powerful characterization of the skeleton of the global system of tax evasion, that is, the strongly connected core of countries (the so called rich-club) that functions as the deep structure of the tax evasion system unveiled by the PPD. Our analysis therefore allows to organize the wealth of available data by means of a clear criterion, that may be conducive to further, useful data mining of the PPD and to a better understanding of the underlying features and properties of the tax evasion network.

2. *Data and methodology*

On 9th May 2016, the International Consortium of Investigative Journalists (ICIJ) released their Offshore Leaks database, including detailed information about more than 300,000 offshore companies and trusts between 1970 and 2016, most of which were registered by the law firm Mossack Fonseca, based in Panama. A task force of hundreds of journalists worldwide has curated the database. The data is one of the largest, up to date, public releases (ICIJ, 2016) of information about offshore entities and lists individuals or companies playing a role in an offshore entity (i.e. "officers") or who help clients to set up an offshore activity (i.e. "intermediaries"). The resulting structure is a tangled web of relationships between individuals and companies based in different countries. Each entity, being either an individual or a company, may be represented as a node in an interacting network, where two nodes are linked if they have a relationship (Fig. 1-A). In fact, each entity is associated to one or more countries. For instance, an individual of country X might be shareholder of a company linked to country Y and registered into another jurisdiction Z.

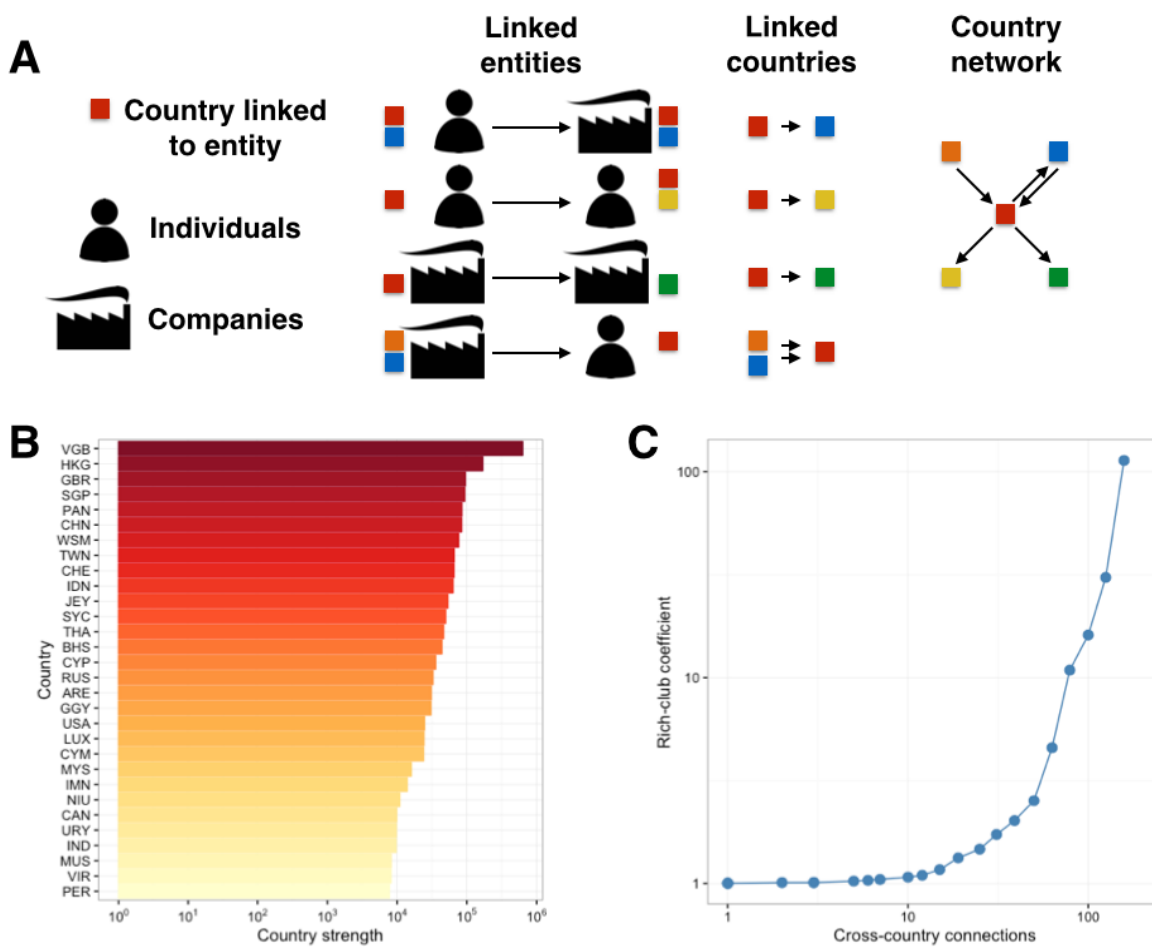

**Figure 1**. *The web of countries hidden in Panama Papers. (A) Entities, either individual or companies, are connected each other by their offshore activities; each entity is linked to one or more countries which in turn establish an effective network of tax evasion. (B) Top 30 countries ranked by the magnitude of their in-coming and out-going activity flow (Country strength). (C) The significant evidence for the existence of a core of highly connected top-ranked countries*

*forming a "rich-club".*

This creates an indirect network of countries, where country X is directly linked to countries Y and Z because of the existence of that shareholder (Fig. 1-A).
Therefore, it is natural to generate a complex network representation (Newman, 2003; Boccaletti et al, 2006) of the Panama Papers information where nodes are countries and links encode their relationships because of shared entities. Connections in this network are weighted (Barrat et al, 2004) by the magnitude of the activity flow: the larger the number of entities bridging two countries, the larger the weight of the link between those countries.

3. Results

As already found in previous studies cited above, the resulting network of countries does not exhibit a trivial structure, and the magnitude of the flow plays a crucial role in structuring the hidden web of tax havens. We probed the network by looking at whether the countries with a high number of connections (i.e. a high degree) tend to preferentially connect to other countries with similar or higher degree. In practice, given the weighted network of relationships, we accounted for the total strength of each country, defined by the sum of the magnitudes of all in-coming and out-going flows. In Fig. 1-B we show the top 30 countries ranked by their total strength. This type of analysis is generally used in network science to reveal the existence of a "rich-club" of nodes (Colizza et al, 2006; Opsahl et al, 2008). The rich-club estimator measures the number of connections among countries with degree higher than a certain threshold with respect to the maximum possible number of connections that they might share. For instance, if the four countries with the highest degree share all possible connections among them, the value of this estimator will be one, the maximum value allowed. Conversely, if those countries are poorly connected, the value of this estimator will approach zero. Measuring the rich-club provides an operational way to identify special correlations in networks among the most central nodes, and allows a better assessment of the robustness of the underlying system against disruption.

To understand whether the observed rich-club effect is a genuine signal of the established tax evasion network, we generate an ensemble of null models that preserves the node's degree distribution of the original country network and the distribution of the connection weights among them. Thus, we calculate the expected rich-club estimator and we compare it against the observed one. The rich-club coefficient, obtained as the ratio between the observed value and the one expected by chance, is a suitable indicator of the rich-club effect (Colizza et al, 2006). When this coefficient is close to 1, the observed effect is due to chance; when it is larger than 1 there is a rich-club effect; when it is smaller than one it indicates the lack of correlations between most central nodes in the network.

Our results for the network obtained from PPD are shown in Fig. 1-C: except for small values of the degrees, there is a strong indication of a rich-club, whose trend is increasing for increasing degree values. For each value of the degree k, there is a corresponding sub-network of countries that are considered the rich-club at the core of the system. While there is no general consensus on which value of k should be considered optimal to extract the network core, our analysis reveals that the rich-club is not altered significantly by the choice of a particular degree value above k = 80. Therefore, we consider this value to extract the core of the network and we report it in Fig. 2.

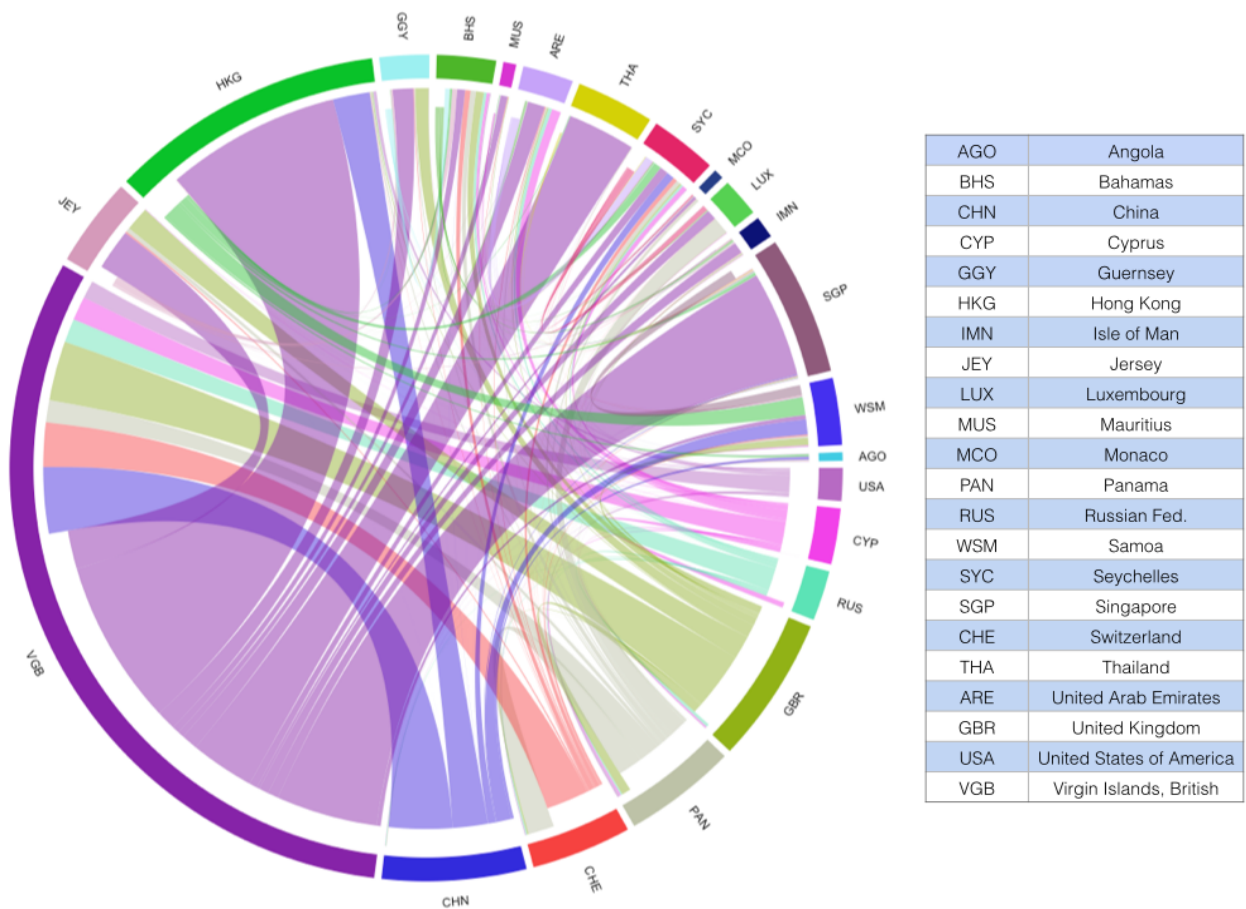

**Figure 2** *The richs' core of tax-evasion country network. Countries in the rich-club, a subset of the whole network, are linked to each other by offshore activities of individuals and companies. In this visualization, wider the ribbon larger the magnitude of the flow between two countries.*

In this context, the observed increasing trend in Fig. 1-C is a strong indicator of the existence of an oligarchy of countries dominating the network. Fig. 2 unveils the identity of countries in the rich-club. They are mainly well-known tax havens, such as Virgin British Islands, Hong Kong and Luxembourg, with the additional involvement of countries like China, Russia, United Kingdom and USA, that is, major powerhouses in the world economy. In agreement with the literature on global tax evasion networks, this result suggests that tax havens and major global economies are structurally intertwined. Channels of tax evasion and avoidance are not exceptional and contingent, but rather systematic and strategic. Dismantling the tax evasion network would likely cause an enormous impact not only on global financial flows, but on the whole organization of the world economy. This is most likely one of the reasons why all attempts to disrupt tax evasion pathways have been largely unsuccessful, despite the availability of increasing, and increasingly detailed, evidence on the structure of such networks and on the identities of the beneficiaries.

Our methodology allows ranking tax havens by their role in the network in terms of strength, i.e. the sum of the magnitudes of incoming and out-going flows. An analysis of such network strengths offers several useful insights into the structure of the rich-club. Interestingly, the rich-club contains some G7 economies, but only a few of them: USA and UK (plus Russia that quit the original G8 group). This does not mean that other G7 countries do not have a role in the global network exposed by PPD (they do), but they do not belong to the network core. This may reflect the combined action of multiple factors: the higher tendency of liberal market economies such as UK or the US to exploit the opportunity offered by tax havens with respect to coordinated market economies such as Germany, Japan, or France (Jones and Temouri, 2016), the role of tax morale (Kemme et al, 2020), and the preferential connections with tax havens that were previous overseas territories for a country's past colonial empire (Palan et al, 2013). Former colonial empires now characterized by a liberal, low tax morale orientation such as the UK are especially likely to play a central role in the global tax evasion network (as well as in other complex global socio-political networks; Erspamer et al, 2022), and not incidentally former British colonies and overseas territories play a central role in the rich-club. On the other hand, former colonial empires with a low tax morale but a coordinated market economy like France are more peripheral. Finally, the foremost non-Western economic power, namely China, does belong to the core, and this is of particular interest in view of its increasingly central role in the new, emerging multipolar world order. Rather than the group of the most socio-economically developed countries as represented in the G7, the rich-club depicted by PPD highlights an East-West axis that paradoxically reflects the new cold war logic: USA and UK on the one side, and China and Russia on the other. It is worth noticing that the latter two countries, despite being authoritarian, have developed a market culture that is closer to that of liberal Western countries than of coordinated Western ones (Tsai, 2011; Light, 2013). However, albeit these countries largely represent competing forces in the multipolar order, they are part of the same cohesive network when it comes to tax avoidance and sheltered financial flows, suggesting that there is a global elite of ultra-rich individuals and multinational enterprises that share a level of concealed agency that bypasses, and even overcomes, nationality, ideological orientations, and even geopolitical interests (Bernstein, 2019).

Despite that tax havens are not limited to micro-states and small islands, in the rich-club emerging from PPD a large majority of members are indeed of this kind. In particular, and again consistently with previous results from the literature, the British Virgin Islands, which generally is not top-ranked according to its secrecy and the scale of its offshore financial activities (Financial Secrecy Index, 2016), has by far the largest role in the rich-club. It should also be noted that the Virgin Islands sit at the top spot in the global ranking of corporate tax havens as of 2019 (Tax Justice Network, 2019). Citizenship in the Virgin Islands is not defined by place of birth but by descent, and this has created a deeply seeded local elite whose privileges have co-evolved with the positioning of the country as a tax haven (Maurer, 1995).

The other micro-states and small islands play a much smaller role in the rich-club, and can be seen as second-tier hubs in the overall architecture; they span a large part of the overall rich-club membership. This result suggests that, should the Virgin Islands tax haven be disrupted for some reason, there would likely be a major reorganization across the remaining tax havens so that another one that now plays a more minor role would become the new pivot. For the typical micro-state tax haven, which is generally affluent but lacking natural resources, taking part in the network is one of the most practical ways to secure high levels of socio-economic development (Dainoff, 2021) – a strategy that, however, may assume a parasitic character from a global economic perspective (Haines, 2017). Dismantling tax evasion networks is therefore not only about breaching secrecy and sanctioning transgressors, but also about providing alternative, viable development models for micro-states with a limited menu of developmental options but high standards of living. Some of the micro-states in the network, however, are such if considered in terms of territorial extension, but are important countries in terms of economic size. This is primarily true for Hong Kong and Singapore, which tellingly play a major organizational role in the rich-club network. Hong Kong is the second most important node in the rich-club after the Virgin Islands, and its prominence is largely related to its close connection with China. Also Singapore features prominently in the rich-club. Other very affluent, geographically larger countries such as Switzerland, the United Arab Emirates and Luxembourg have a lesser but still important role.

Not all countries in the rich-club are actually rich. The network also includes poor (but resource rich) countries such as Angola, and emerging countries such as Thailand. Other countries such as Cyprus (which, together with Luxembourg is the only EU member state in the rich-club) has quickly transitioned from the emerging to the affluent tier of world economies as its income per capita has more than tripled in the last thirty years, most likely also thanks to its gateway role for sheltered Russian foreign direct investment and money laundering (Repousis et al, 2019).

Overall, then, the rich-club of the global tax evasion network is a complex mix of heterogeneous members: Western and Eastern global powers and micro-states, rich and poor countries, resource-rich and resource-poor ones, educationally and technologically advanced and underdeveloped ones, which, as noted by Haberly and Wójcik (2015) in the context of global offshore FDI networks, reflects several historical layers of socio-economic change, from the demise of the UK colonial empire to the rise of China as a new economic superpower. In view of such heterogeneity, it is clear that no specific set of socio-economic characteristics is what makes the network cohesive. The assembly of countries that belong to the rich-club is apparently kept together by their complementarity in securing the smooth and flexible functioning of a tax evasion scheme that is able to shelter flows from large and/or rich countries and ultra-rich individuals to convenient, highly interconnected repositories of assets whose ownership and origin are completely concealed and not traceable. As new powers have risen to the global scene like in the case of China, they have apparently been co-opted into the rich-club. In such network, criminal entities have unique opportunities to infiltrate legal ones under the shield of secrecy and anonymity (Passas, 2003), critically undermining the civil constituency of the world economy.

*4. Conclusions*

There is no simple solution to dismantling the global networks of tax evasion. As already remarked, fighting tax havens by closing down a sub-set of them would reduce the competition among the ones that remain active, thus enhancing their opportunities (Elsayyad and Konrad, 2012) and possibly their centrality in the re-shaped rich-set. The overall picture might be even worse. In fact, our results show that the phenomenon of offshore activity cannot be completely understood if tax havens are treated as independent units, instead calling for a network perspective, where tax havens represent the core of a much larger worldwide web of offshore activities, so that effective disruption is made complex by the fact that in all countries there are powerful interest groups that benefit from its existence. Overcoming internal resistance and pressure by the local, often invisible vested interests is especially difficult, and this may undermine both legitimacy and effectiveness of anti-tax evasion policies.

The understanding of the structure of the skeleton of tax havens, and the identification and ranking of the countries that are most closely involved in the shielding of wealth in a networked context provides a measure of the difficulties that target policies will have to face to overcome tax evasion. For the purpose of policy design, the rich-club should not be regarded as a static structure to intervene upon, but rather as a flexible mechanism that can quickly reorganize as a consequence of any targeted disruptive intervention, so that optimal policy action should be designed to effectively address the system's dynamic response rather than its current organization.

In this perspective, and in view of the ineffectiveness of partial disruptive interventions that would likely only lead to a redistribution of resources and centrality across the still extant havens, the idea of tackling global tax evasion not through attempts at eradication but through gradual, systematic integration of tax havens into the global tax network in the context of a broader social sustainability strategy (Bird and Davis-Nozemack, 2018) is possibly both more feasible and more effective. However, the timing and coordination of such integration processes is critical, and a detailed simulation analysis may be extremely helpful to fine tune (quasi-)optimal control strategies in the implementation phase.

Modelling possible dynamic response scenarios to given policy interventions is a challenging analytical task that needs to be urgently addressed by future research. Computational social science may play a crucial role in this ambitious and necessary endeavor, which must leverage an unprecedented interdisciplinary collaboration with geography, economics, international law, political science and security studies. Without proper analytical tools, and in view of the serious frictions against effective intervention, the goal of developing a globally integrated, transparent tax system, which would amount to a crucial step on the way to global fiscal justice, would not only be difficult, but unattainable in principle.